\documentclass[12pt,draftcls,onecolumn]{IEEEtran}
\usepackage{amssymb,bbm,amsmath,color,amsthm,psfrag,subfigure}
\usepackage[draft]{graphicx}
\usepackage{xspace}
\usepackage{algorithm,algorithmic} 
\usepackage{wrapfig}
\usepackage{tikz}

\def\beq{\begin{equation}}
\def\eeq{\end{equation}}

\def\beq{\begin{equation}}
\def\eeq{\end{equation}}

\newtheorem{theorem}{Theorem}

\newtheorem{proposition}{Proposition}

\theoremstyle{remark}

\newcommand{\ds}{\displaystyle}

\newcommand{\ba}{\begin{array}}
\newcommand{\ea}{\end{array}}

\renewcommand{\l}{\left}\renewcommand{\r}{\right}
\newcommand{\be}{\begin{equation}}
\newcommand{\ee}{\end{equation}}

\newcommand{\1}{\mathbbm{1}}

\newcommand{\R}{\mathbb{R}}

\newcommand{\de}{\mathrm{d}}

\newcommand{\subscr}[2]{{#1}_{\textup{#2}}}

\newcommand{\setdef}[2]{\{#1 \; | \; #2\}}

\newcommand{\mc}{\mathcal}

\newcommand{\flow}{f}

\newcommand{\densitymaxe}{\rho^{\text{max}}_{e}}

\newcommand{\dist}{\mathrm{dist}}
\newcommand{\newflowmaxe}{\subscr{\tilde{f}}{e}^\textup{max}}
\DeclareMathOperator{\cl}{cl}

\newcommand{\nodeset}{\mc V}
\newcommand{\edgeset}{\mc E}

\newcommand{\flowmaxe}{\subscr{f}{e}^\textup{max}}

  \renewenvironment{thebibliography}[1]{%
    \begin{oldthebibliography}{#1}%
      \setlength{\parskip}{0.02ex}%
      \setlength{\itemsep}{0.02ex}%
  }%
  {%
    \end{oldthebibliography}%
  }

\newcommand\oprocendsymbol{\hbox{$\square$}}
\newcommand\oprocend{\relax\ifmmode\else\unskip\hfill\fi\oprocendsymbol}

\title{Robust Distributed Routing in Dynamical Networks with Cascading Failures \thanks{This work was supported in part by NSF EFRI-ARES grant number 0735956 and AFOSR grant number FA9550-09-1-0538.}}

\author{Giacomo Como \thanks{Department of Automatic Control, Lund University, Sweden. \texttt{giacomo.como@control.lth.se}} \quad Ketan Savla \thanks{Laboratory for Information and Decision Systems, Massachusetts Institute of Technology, Cambridge, MA, USA. \texttt{ksavla@mit.edu}.} \quad Daron Acemoglu
\thanks{Department of Economics,
Massachusetts Institute of Technology, Cambridge, MA, USA.
({\tt daron@mit.edu})}
\quad Munther A.~Dahleh
\thanks{Laboratory for Information and Decision Systems,
Massachusetts Institute of Technology, Cambridge, MA, USA.
({\tt dahleh@mit.edu})}
\quad Emilio Frazzoli 
\thanks{Laboratory for Information and Decision Systems,
Massachusetts Institute of Technology, Cambridge, MA, USA.
({\tt frazzoli@mit.edu})} }

\begin{document}

\maketitle



\vspace{-0.1in}
\section{Overview} Robustness of routing policies for networks is a central problem which is gaining increased attention with a growing awareness to safeguard critical infrastructure networks against natural and man-induced disruptions. Routing under limited information and the possibility of cascades through the network adds serious challenges to this problem. 
This abstract considers the framework of \emph{dynamical networks} introduced in our earlier work~\cite{Como.Savla.ea:Part1TAC10,Como.Savla.ea:Part2TAC10}, where the network is modeled by a system of ordinary differential equations derived from mass conservation laws on directed acyclic graphs with a single origin-destination pair and a constant inflow at the origin. The rate of change of the particle density on each link of the network equals the difference between the \emph{inflow} and the \emph{outflow} on that link. The latter is modeled to depend on the current particle density on that link through a \emph{flow function}. The novel modeling element in this abstract is that every link is assumed to have finite capacity for particle density and that the flow function is modeled to be strictly increasing as density increases from zero up to the maximum density capacity, and is discontinuous at the maximum density capacity, with the flow function value being zero at that point. This feature, in particular, allows for the possibility of spill-backs in our model, which was absent from the model considered in \cite{Como.Savla.ea:Part1TAC10,Como.Savla.ea:Part2TAC10}. In this extended abstract, we present our results on resilience of such networks under distributed routing, towards perturbations that reduce link-wise flow functions. 

\section{Model formulation} For the sake of this abstract, we consider the following simple scenario.
The topology of the network is abstracted as an acyclic directed graph $\mc T=(\nodeset,\edgeset)$ with one origin/destination pair and such that every node lies on a path from the origin to the destination. If $e=(v,w)\in\mc E$ is a link, we shall write $\sigma(e)=v$ and $\tau(e)=w$  for its tail and head node, respectively.
Acyclicity implies that we can identify the node set $\mc V=\{0,1,\ldots,n\}$ in such a way that $\sigma(e)<\tau(e)$ for all $e\in\mc E$, so that in particular $0$ is the origin and $n$ the destination. The sets of outgoing  and incoming links of a node $v\in\mc V$ will be denoted by $\mc E^+_v:=\{e\in\mc E:\,\sigma(e)=v\}$ and $\mc E^-_v:=\{e\in\mc E:\,\tau(e)=v\}$, respectively. The state of the network is described by a vector $\rho=\{\rho_e:\,e\in\mc E\}$, where $\rho_e$ denotes the density on link $e$. 
The flow on link $e$ is given by the functional relationship 
$f_e=\mu_e(\rho_e)$. We assume that the flow function $\mu_e:[0,\densitymaxe]\to\R^+$ is continuous differentiable and strictly increasing on $[0,\densitymaxe)$, and is such that $\mu_e(0)=\mu_e(\densitymaxe)=0$. Here, $\densitymaxe>0$ stands for the density capacity on link $e$. We let $\Gamma:=\times_{e\in\mc E}[0,\densitymaxe)$ be the state space, and the maximum flow capacities be
$\flowmaxe:=\lim_{\rho_e \uparrow \densitymaxe} \mu_e(\rho_e)$ for all $e \in \mc E$. We shall write $\mu:=\{\mu_e:e\in\mc E\}$ for the vector of flow functions. If $\tilde\mu:=\{\tilde\mu_e:e\in\mc E\}$ is another flow function vector with the same properties, an inequality $\mu\ge\tilde\mu$ will have to be interpreted component-wise. The way the inflow of a node $v<n$ gets split among its outgoing links is determined by the routing policy $\mc G:=\{G^v:\Gamma_v\to\mc S_v\}_{0\le v<n}$, where $\Gamma_v:=\times_{e\in\mc E^+_v}[0,\densitymaxe)$, and $\mc S_v$ stands for the simplex of probability vectors over $\mc E^+_v$. In particular, we focus on the class of \emph{locally responsive} distributed routing policies, which is a family of continuously differentiable distributed routing functions $\mc G=\{G^v:\Gamma_v\to\mc S_v\}_{v\in\mc V}$ such that, for every non-destination node $0\le v<n$:
\begin{description}

\item[(a)]
for every nonempty proper subset $\mc J\subsetneq\mc E^+_v$, there exists a continuously differentiable map $G^{\mc J}: \Gamma_{\mc J}\to\mc S_{\mc J}$, where $\mc R_{\mc J}:=\R_+^{\mc J}$, $\Gamma_{\mc J}:=\setdef{\rho \in \mc R_{\mc J}}{\rho_e < \densitymaxe \quad \forall e \in \mc J}$, and $\mc S_{\mc J}:=\{p\in\mc R_{\mc J}:\,\sum_{j\in\mc J}p_j=1\}$ is the simplex of probability vectors over $\mc J$, such that, for every $x^{\mc J}\in\Gamma_{\mc J}$, if $$x_e \to \densitymaxe\,,\ \ \forall e\in\mc E^+_v\setminus\mc J\,, x_j\to x_j^{\mc J} \in \Gamma_{\mc J}\,,\ \ \forall j\in\mc J\,, \quad \text{then}$$ \vspace{-0.05in}
$$G^v_e(x)  \to0,\ \ \forall e\in\mc E^+_v\setminus\mc J\,, \quad G^v_j(x)  \to G^{\mc J}_j(x^{\mc J}),\ \ \forall j\in\mc J\,.
$$ \vspace{-0.1in}
\item[(b)]
$\ds\frac{\partial}{\partial x_e}G^v_j(x)\ge0 \,, \forall j,e\in\mc E^+_v\,,  j\ne e\,,x \in \Gamma_v\,,$
\end{description}
 
The two salient features of this routing policy are the \emph{local information constraint} which allows the routing policy $G^v(\rho^v)$ to depend only on the particle density $\rho^v$ on the set $\mc E^+_v$ of outgoing links of the non-destination node $v$, and the condition (a), which models the fact that no flow can be routed to a fully congested link.
The additional condition (b) is a rather natural in that it states that the faction of particles routed towards any link does not decrease when the particle density in some other link is increased. 
It is reminiscent of the notion of cooperative dynamical system. 
In fact, routing policies with this property were proven to be optimal in terms of resilience in the our earlier work on dynamical networks with infinite density capacity \cite{Como.Savla.ea:Part1TAC10,Como.Savla.ea:Part2TAC10}.  
 
We are now ready to describe the dynamical network model used in this abstract. Consider a flow network over $\mc T$ with a constant inflow $\lambda_0\ge0$ at the origin and the flows on the links evolving according to the following dynamical system with $\rho(t) \in \cl(\Gamma)$ as the state vector:
\begin{equation}
\label{dynsyst}
\frac{\de}{\de t}{\rho}_e(t)  =\chi_{\sigma(e)}(t)\lambda_{\sigma(e)}(t)G^{\sigma(e)}_e\l(\rho^{\sigma(e)}(t)\r)  -\chi_{\tau(e)}(t) f_e(t)\,,
\end{equation}
for all $e\in\mc E$, where $G_v$ are the locally responsive routing policies, $f(t)=\mu(\rho(t))$ and $\lambda_v(t)$ is the incoming flow at node $v \in \mc V$ and is equal to $\lambda_0$ if $v=0$ and equal to $\sum_{e\in\mc E^-_v}f_e(t)$ if $v>0$, while
$\chi_v(t):=1-\prod_{e\in\mc E^+_v}(1-\xi_e(t))$ and $\xi_e(t):=\1_{[0,\densitymaxe)}(\rho_e(t))$
are the activation status indicators of a node $v\in\mc V$ 
and a link $e\in\mc E$. In this abstract, we use \emph{dynamical network} to  refer to the dynamics (\ref{dynsyst}) with locally responsive routing policies $\mc G$.

Equation (\ref{dynsyst}) states that the rate of change of the particle density on a link $e$ outgoing from some non-destination node $v$ is given by the difference between $\lambda_v(t)G^v_e(\rho^v(t))$, i.e., the portion of the total outflow at node $v$ which is routed to link $e$, and $f_e(t)$, i.e., the particle flow on link $e$. The set of such equations model conservation of mass both at every non-destination node and on the links of the flow network. In particular, when $\chi_v(t)=0$, no flow can be absorbed by any of the outgoing links of node $v$, and (\ref{dynsyst}) implies that no flow comes out of any of the incoming links of node $v$. Observe that the distributed routing policy $G^v(\rho^v)$ induces a local feedback which couples the dynamics of the particle flow on the different links. In fact, the dynamical network (\ref{dynsyst}) should be interpreted as an $|\mc E|$-dimensional switched system.  Existence and uniqueness of a solution for every initial density $\rho(0)\in  \cl(\Gamma)$,  then follow from the differentiability assumptions on the flow function $\mu$ and the routing policy $\mc G$ by standard arguments. 

The most novel feature of the dynamics (\ref{dynsyst}) resides in the role of the link and node activation status indicators $\xi_e(t)$, and $\chi_v(t)$. Indeed, observe that, if $\xi_e(t^*)=0$ for some $t^*$, then $\xi_e(t)=0$ for all $t\ge t^*$. This is a direct consequence of the fact that $\lambda_{\sigma(e)}(t)G^v_e(\rho^v)-\mu_e(\rho_e)=0$ whenever $\rho_e=\densitymaxe$. Once the particle density reaches its maximum capacity on a link, the corresponding outflow is zero, and the link becomes irreversibly inactive. On the other hand, the definition of indicator variables implies that a node becomes inactive, or fails, when all the outgoing links do so, and thus it remains inactive ever since. In turn, this drops the outflow of all its incoming links to zero so that they are bound to become inactive. As a consequence some other links may experience an overload, possibly reaching their density capacity, thus becoming inactive ever since. Through this mechanism, link and node failures can propagate through the network. 

The following proposition states a fundamental dichotomy in the behavior of the dynamical network: either all the asymptotic time-averaged outflow equals the constant inflow, or it is zero. Such a dichotomy is a direct consequence of the boundedness of the density capacities.

\begin{proposition} \label{properties}
For any initial density vector $\rho(0)\in \cl(\Gamma) $, either of the following alternatives hold:
\be\label{alternative1}
\lim_{t\to\infty}\frac1t\int_{0}^t\lambda_n(s)\de s=\lambda_0\,, \quad \text{or} \quad \lim_{t\to\infty}\frac1t\int_{0}^t\lambda_n(s)\de s=0.
\ee
\end{proposition}

Motivated by this, we define the dynamical network to be  \emph{transferring} with respect to some initial density vector $\rho(0)\in\cl (\Gamma)$ if the first condition in (\ref{alternative1}) holds. We consider persistent perturbations of the dynamical network (\ref{dynsyst}) that reduce the flow functions on the links. Formally, an \emph{admissible perturbation} of a dynamical network is a network with the same topology $\mc T$, same locally responsive distributed routing policies $\mc G=\{G^v:\Gamma_v\to\mc S_v\}_{v\in\mc V}$ and a family of perturbed flow functions $\tilde\mu:=\{\tilde\mu_e: [0,\rho_e^{\text{max}}] \to\R_+\}_{e\in\mc E}$ that have the same monotonicity and differentiable properties as $\mu$, and are such that, for every $e\in\mc E$, $\tilde\mu_e(\rho_e)\le\mu_e(\rho_e)$, for all $\rho_e\in[0,\densitymaxe]$. We accordingly let $\newflowmaxe:=\lim_{\rho_e \uparrow \densitymaxe} \tilde{\mu}_e(\rho_e)$. The \emph{perturbed dynamical network} is then governed by 
\vspace{-0.05in}
\begin{equation}
\label{dynsyst-perturbed}
\frac{\de}{\de t}{\tilde{\rho}}_e(t)  =\tilde{\chi}_{\sigma(e)}(t)\tilde{\lambda}_{\sigma(e)}(t)G^{\sigma(e)}_e\l(\tilde{\rho}^{\sigma(e)}(t)\r)  -\tilde{\chi}_{\tau(e)}(t) \tilde{f}_e(t)\,,
\end{equation}
for all $e\in\mc E$, where $\tilde{f}(t)=\tilde{\mu}(\tilde{\rho}(t))$ and $\tilde{\lambda}_v(t)$ is equal to $\lambda_0$ if $v=0$ and equal to $\sum_{e\in\mc E^-_v}\tilde{f}_e(t)$ if $v>0$, while
$\tilde{\chi}_v(t):=1-\prod_{e\in\mc E^+_v}(1-\tilde{\xi}_e(t))$ and $\tilde{\xi}_e(t):=\1_{[0,\densitymaxe)}(\tilde{\rho}_e(t))$ for all $v \in \mc V$. Note that the routing policy $\mc G$ used in \eqref{dynsyst} and \eqref{dynsyst-perturbed} are the same.

The \emph{magnitude} of an admissible perturbation is defined as 
$||\delta||_1=\sum_{e \in \mc E}\delta_e\,,$ where
$\delta\in\R_+^{\mc E}\,, \delta_e:=\sup_{\rho_e \in [0,\rho_e^{\text{max}}]}\l\{\mu_e(\rho_e)-\tilde\mu_e(\rho_e)\r\}\,,\ e\in\mc E\,.$
The \emph{margin of resilience} of the dynamical network $\gamma(\rho^{\circ})$ is defined as the infimum magnitude of all the admissible perturbations for which the perturbed dynamical network is not transferring with respect to the initial density vector $\tilde\rho(0)=\rho^{\circ}$.  
In the next section, we report our results on the margin of resilience.

\section{Results} 
We start by stating an upper bound on the margin of resilience of a dynamical network. 
Throughout, we shall assume that $\rho^{\circ}$ is an equilibrium for the unperturbed dynamical network\footnote{The conditions for existence and uniqueness of equilibrium follows from our prior work~\cite{Como.Savla.ea:Part1TAC10} on dynamical flow networks with unbounded density capacities.}, with the corresponding flow $f^{\circ}=\mu(\rho^{\circ})$, i.e., $\lambda_0 G^0_e(\rho^{\circ})=f_e^{\circ}$ for all $e \in \mc E_0^+$ and $(\sum_{e \in \mc E_v^-} f_e^{\circ}) G^v_e(\rho^{\circ})=f_e^{\circ}$ for all $e \in \mc E_v^+$ when $v>0$.
We shall also assume that the topology $\mc T$ is tree-like. By this, we mean that the only node reachable from the origin by two distinct paths is the destination one. The assumption of tree-likeness of the topology implies that one can partition the node set as 
$\mc V=\bigcup_{0\le j\le j^*}\mc V_j\cup\{n\}\,,$
where $\mc V_j:=\{0\le v<n:\,\dist(0,v)=j\}$ is the set of non-destination nodes at distance $j$ from the origin, and $j^*$ is the maximal distance of a non-destination node from the origin.

Before proceeding we introduce here some preliminary notation. 
Let us introduce the residual capacity  
$R_v(f^{\circ}):=\sum_{e\in\mc E^+_v}\flowmaxe-\flow_e^{\circ}\,,$
of a non-destination node $0\le v<n$, and let 
$R(f^{\circ}):=\min\{R_v(f^{\circ}):\,0\le v<n\}$
be the minimal node residual capacity of the network. 
For an inflow $\lambda_0\ge0$, we shall consider the set of equilibrium flows 
$$\mc F(\lambda_0):=\l\{f \in\mc F:\,\sum\nolimits_{e\in\mc E^+_0}f_e=\lambda_0\,,
\sum\nolimits_{e\in\mc E^+_v}f_e=\sum\nolimits_{e\in\mc E^-_v}f_e\,,\quad \forall \, 0<v<n\r\}\,.$$
An origin-destination cut is a subset $\mc U\subseteq\mc V$ such that $0\in\mc U$ and $n\notin\mc U$. For an origin-destination cut $\mc U$, let $\mc E^+_{\mc U}:=\{e\in\mc E:\,\sigma(e)\in\mc U,\tau(e)\notin\mc U\}$ be the set of links with tail node in $\mc U$ and head node in $\mc V\setminus\mc U$, and let 
$\mc C(\mc U):=\sum\nolimits_{e\in\mc E^+_{\mc U}}\flowmaxe$ be its capacity. The \emph{min-cut capacity} of the flow network is 
$\mc C :=\min_{\mc U} C(\mc U)\,,$
where the minimization runs over all the origin-destination cuts. By the min-cut max-flow theorem it follows that $\mc F(\lambda_0)\ne\emptyset$ if and only if $C(\mc N)\ge\lambda_0$, a condition that we shall assume in order to avoid trivialities.

\textbf{(a)}  We now describe a procedure to compute an upper bound on the margin of resilience. 
For every non-destination node $0\le v<n$, and $\lambda\ge0$, define 
$$\mc X_{v}(\lambda):=\l\{x\in\times_{e}[0,\flowmaxe]:\,\sum\nolimits_{e}\l(\flowmaxe-x_e\r)\le\lambda \r\}\,.$$
where the product/sum index $e$ runs over $\mc E^+_v$.
Further, let $d_n=+\infty$.  For $j=j^*,j^*-1,\ldots,0$, iteratively define 
\begin{equation*}
\label{dvdef}
d_{v}:=\min\l\{c_v(x):\,x\in\mc X_v(\lambda^{\circ}_v)\r\}\,,\qquad\lambda^{\circ}_v:=\sum\nolimits_{e\in\mc E^+_v}f_e^{\circ}\,,
\end{equation*}
for all $v\in\mc V_j$, where
$c_v(x):=\sum\nolimits_{e\in\mc E^+_v}\min\{x_e,d_{\tau(e)}\}\,.$
The intuition behind this definition is the following: $c_v(x)$ is the cost that a hypothetical malicious adversary has to face in order to reduce the sum of the maximal flow capacities of the outgoing links of a node $v$ below the inflow $\lambda_v^{\circ}$, thus causing the eventual link's failure. In order to compute such cost, for every outgoing link $e$, the minimum between the flow capacity reduction $x_e$ and the previously computed cost to induce a failure of the head node $\tau(e)$ is considered. 

The following result states an upper bound on the margin of resilience in terms of the calculations described above.

\begin{theorem}[Upper bound on the margin of resilience]
\label{thm:resilience-upper-bound}
Assume that the network topology $\mc T$ is tree-like, and that the dynamical network has an equilibrium density vector $\rho^{\circ}$. 
Then, the margin of the resilience is upper bounded as:
$\gamma(\rho^{\circ}) \le d_0\,.$
\end{theorem}

%
The following proposition gives useful bounds on the the quantity $d_0$.
\begin{proposition}
\label{prop:d0-bounds}
Let 
$f^{\circ}=\mu(\rho^{\circ}) \in \mc F(\lambda_0)$ be an equilibrium flow for the dynamical network. Then, 
$R(f^{\circ})\le d_0\le \mc C-\lambda_0\,.$
\end{proposition}

Proposition~\ref{prop:d0-bounds} and Theorem~\ref{thm:resilience-upper-bound} imply that the margin of resilience is upper bounded by the network residual capacity $\mc C-\lambda_0$. In fact, one can show examples where the margin of resilience is strictly less than $\mc C-\lambda_0$.
In \cite{Como.Savla.ea:Part1TAC10}, we showed that the margin of (weak) resilience is equal to $\mc C$ when there is no bound on the density capacity. This illustrates the loss in resilience due to finiteness of density capacities. 

\medskip

\textbf{(b)} In \cite{Como.Savla.ea:Part2TAC10}, we computed margin of \emph{strong} resilience of a dynamical network when there is no bound on the maximum particle density on the links, and the flow functions are monotonically increasing. The margin of strong resilience is defined to be the infimum magnitude of all admissible perturbations for which (analogous to the definition of margin of resilience in this abstract) the outflow from the destination node of the perturbed dynamical network is not asymptotically equal to $\lambda_0$. In particular,  
we showed that the margin of strong resilience in that setting is equal to the minimum node residual capacity of the network $R(f^{\circ})$. We show that, when the links have finite capacity for particle densities as in this abstract, the margin of resilience could be possibly \emph{greater} than the minimum node residual capacity. We illustrate this point through the following example. 

\begin{wrapfigure}{l}{85mm}
\vspace{-0.2in}
 \begin{center}
\includegraphics[width=6cm,height=3cm]{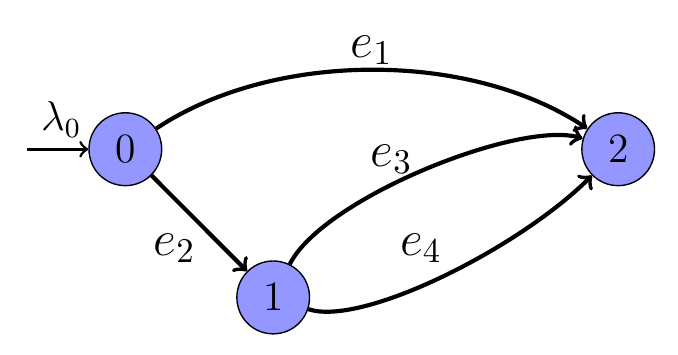}
\end{center}
\vspace{-0.15in}
\caption{\label{fig:A5-justification} A sample network topology.}
 \vspace{-0.1in}
  \end{wrapfigure}
Consider the topology shown in Figure~\ref{fig:A5-justification}. Let the flow functions $\mu$ be such that: 
$f^{\text{max}}_{e_1}=3$, $f^{\text{max}}_{e_2}=1.5$, $f^{\text{max}}_{e_3}=f^{\text{max}}_{e_4}=0.75$. Let the equilibrium flows be: $f^{\circ}_{e_1}=f^{\circ}_{e_2}=1$ and $f^{\circ}_{e_3}=f^{\circ}_{e_4}=0.5$. The min node residual capacity with these parameters is $0.5$. We now show that, if nodes $0$ and $1$ are implementing locally responsive distributed routing policy, then even with a disturbance of magnitude $0.6>0.5$, then network is still transferring.
First, consider a specific disturbance of magnitude 0.6 under which the perturbed flow functions are: $\tilde{\mu}_{e_1}=\mu_{e_1}$, $\tilde{\mu}_{e_2}=\mu_{e_2}$, $\tilde{\mu}_{e_3}=3 \mu_{e_3}/5$ and $\tilde{\mu}_{e_4}=3 \mu_{e_4}/5$. For the perturbed network, $\tilde{f}^{\text{max}}_{e_3}+\tilde{f}^{\text{max}}_{e_4} = 0.9 < 1 = f^{\circ}_{e_3}+f^{\circ}_{e_4}$. Therefore, after a finite time, both $\tilde{\rho}_{e_3}(t)$ and $\tilde{\rho}_{e_4}(t)$ hit the respective maximum capacity on particle densities. at which point $\chi_1$ becomes zero. As a consequence the outflow term for link $e_2$ becomes zero after this time. If the routing policy $G$ at node $0$ has property that $G^0_{e_2}(\tilde{\rho}^v) \to 0$ if $\tilde{\rho}_{e_2} \to \rho^{\text{max}}_{e_2}$, the inflow of $2$ at node $0$ is routed to link $e_1$ and hence the network maintains its transferring property. 
In general, for any other disturbance of magnitude $0.6$, in the worst-case, the inflow to node $1$ would be such that it could exceed the sum of perturbed capacities of links $e_3$ and $e_4$ and hence making $\chi_1=1$, after which one can repeat the argument to show that all the inflow of 2 at node $0$ is transferred to link $e_1$ and the network maintains its transferring property.

This example shows that spill-backs act as backward propagators of information to upstream routing policies (in this example, the \emph{local} routing policy at node $0$ gets information about links $e_3$ and $e_4$ through spill-backs). Since having information about downstream links by routing policies increases resilience, spill-back cascades lead to an increase in the margin of resilience of dynamical networks.

\section{Conclusion} In this paper, we studied resilience of capacitated dynamical networks, where the links have finite capacity for flow density, thereby allowing the possibility of spill-back cascades. We studied the effect of such cascades on the resilience of the network and provided an algorithm to compute an upper bound on the margin of resilience of the network for tree-like topologies. Future work will involve performing analysis for general acyclic and cyclic topologies, possibly with multiple origin-destination pairs.

\thispagestyle{empty}
\vspace{-0.1in}
{\small
  \bibliographystyle{ieeetr}%
  \bibliography{KS-transportation}
}

\end{document}